\documentclass[letter]{aa}
\usepackage{amssymb,amsmath}
\usepackage{graphicx}
\usepackage{xcolor,natbib}
\usepackage{multirow}
\usepackage[T1]{fontenc}
\usepackage{ae,aecompl}
\usepackage{newtxtext, newtxmath}
\usepackage{float}
\usepackage{subfigure}
\usepackage{longtable}
\usepackage{enumitem}
\usepackage{longtable,listings}
\usepackage[flushleft]{threeparttable}
\usepackage{parcolumns}
\usepackage{hyperref}

\def\n2{[N~{\sc ii}]$\lambda6583$\AA}

\def\o3{[O~{\sc iii}]$\lambda4364$\AA}

\begin{document}

\title{More than 60\% of double-peaked narrow emission lines not related to dual galaxy systems?}

\titlerunning{double-peaked narrow lines}

\author{XueGuang Zhang}

\institute{Guangxi Key Laboratory for Relativistic Astrophysics, School of Physical Science and Technology, GuangXi University, 
No. 100, Daxue Road, Nanning, 530004, P. R. China \ \ \ \email{xgzhang@gxu.edu.cn}}

\abstract
{Dual galaxy system (DGS) is one of the widely accepted scenarios to explain the double-peaked narrow emission lines (DPNELs) 
due to orbital motions of the two galaxy in a merging system. After considering no physical connections between two 
independent narrow emission line regions in two galaxies in one DGS, there should be no correlations between flux ratios $R_{R}$ of 
red-shifted narrow emission components from one galaxy and flux ratios $R_{B}$ of blue-shifted narrow emission components from 
the other galaxy in the DGS. However, after checking the large sample of DPNELs in the SDSS, there are strong linear correlations 
in different groups between $R_{R}$ as the flux ratio of red-shifted narrow [O~{\sc iii}] to the red-shifted narrow H$\alpha$ 
and $R_{B}$ as the flux ratio of blue-shifted narrow [O~{\sc iii}] to the blue-shifted narrow H$\alpha$. Meanwhile, after checking 
narrow emission line properties of galaxy pairs within 30 (20, 10) arcmins, there are no connections between narrow emission line 
fluxes in the galaxy pairs, to support the detected linear correlations being robust enough between $R_{R}$ and $R_{B}$ in the 
DPNELs in SDSS. Furthermore, through oversimplified simulations, at least more than 60\% of the DPNELs should be not related to 
the expected DGSs.  
}


\maketitle

\section{Introduction}

	Narrow emission lines from central narrow emission line regions (NLRs) are fundamental optical spectroscopic 
characteristics of emission line galaxies, such as active galactic nuclei (AGN) and HII galaxies. Moreover, among emission line 
galaxies, there is one unique subclass, the emission line galaxies with double-peaked narrow emission lines (DPN galaxies). 
Since the first DPN galaxy reported in \citet{zhou04}, more 10000 DPN galaxies have been reported, such as the samples in 
\citet{cg09, wang09, sm10, ge12, wl19, mm20, zheng25}.

	In order to explain the double-peaked narrow emission lines (DPNELs), the following two main scenarios have been proposed. 
First, two independent NLRs related to two galaxies merging as a dual galaxy system (DGS) can normally lead to DPNELs 
due to orbital motions of the two galaxies, as natural products of hierarchical galaxy formation and evolution \citep{sr98, 
cl00, vh03, md06, db07, se14, zd19, yp22, ag24}. Second, local dynamic structures (such as radial flows, rotating disks) in NLRs 
in individual galaxy can also lead to DPNELs, such as the discussions in \citet{sh11, fy12, nc16, cn18, rd19}.

	It is clear that not all the DPN galaxies can be accepted as indicators of DGSs, but it is not clear that the proportion 
of DPN galaxies are not related to DGSs. Here, an independent method is proposed to determine the proportion of DPN galaxies not 
related to DGSs. Assumed DPNELs related to DGSs, corresponding spatial separation of the two galaxy in one DGS should 
be longer enough that narrow line emission properties were self-governed by each individual galaxy in the DGS, otherwise more 
complicated profiles should be expected rather than apparent double-peaked features. In other words, very weak intrinsic 
connections could be expected between narrow emission lines from the two galaxies in one DGS leading to DPNELs. Therefore, to 
check emission property connections between the red-shifted components and the blue-shifted components in DPNELs will provide 
meaningful clues to support or to be against the DPNELs related to the proposed DGSs, which is our main objective.

	The manuscript is organized as follows. Section 2 presents the main results and necessary discussions on tight flux 
correlations between the red-shifted components and the blue-shifted components in DPNELs. The main summary and conclusions 
are given in Section 3. In the manuscript, we have adopted the cosmological parameters of $H_{0}$=70 km s$^{-1}$ Mpc$^{-1}$, 
$\Omega_{m}$=0.3, and $\Omega_{\Lambda}$=0.7.

\begin{figure*}
\centering\includegraphics[width = 18cm,height=4cm]{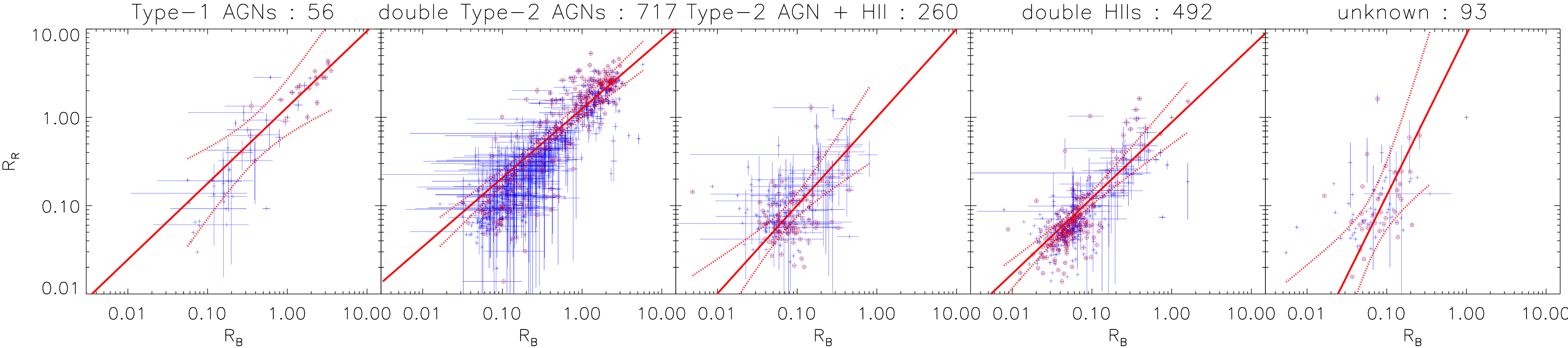}
\caption{On the correlations between $R_R$ and $R_B$ for the SDSS objects with apparent DPNELs in the five groups. In each panel, 
solid red line and dashed red lines show the best fitting results and the corresponding $3\sigma$ confidence bands. In 
each panel, open red circles mark the galaxies with larger $R_{ps}$.}
\label{rat}
\end{figure*}

\begin{figure}
\centering\includegraphics[width = 9cm,height=10cm]{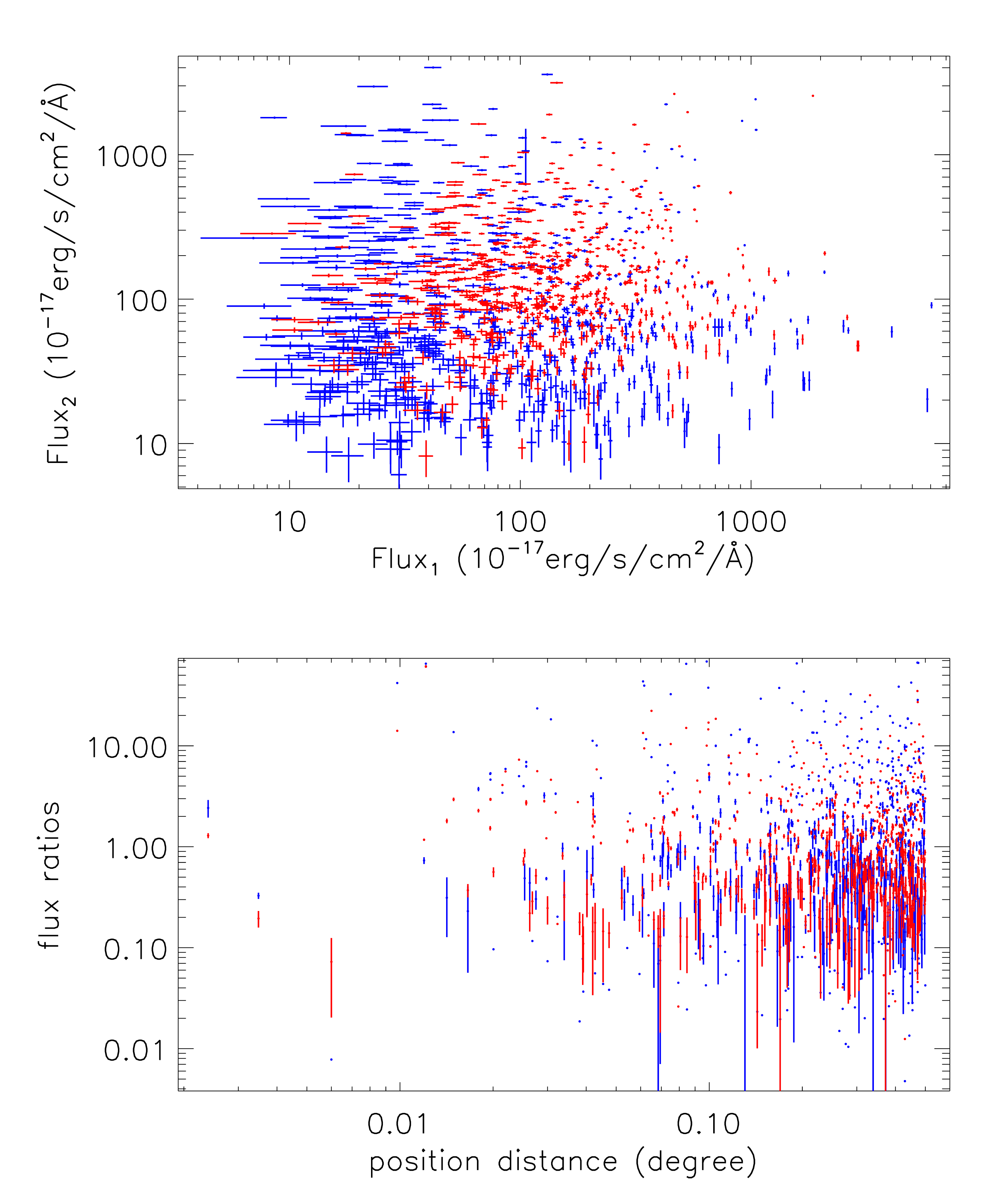}
\caption{Top panel shows the flux correlations of narrow emission lines of both [O~{\sc iii}]$\lambda5007$\AA~ (in blue) and 
narrow H$\alpha$ (in red) in the 669 AGN pairs with redshift difference smaller than 0.001 and with position distance smaller 
than 30 arcmins. Bottom panel shows the dependencies of narrow emission line flux ratios on position distance. In bottom panel, 
symbols in blue and in red show the flux ratios of [O~{\sc iii}]$\lambda5007$\AA~ and the flux ratios of narrow H$\alpha$ in 
the 669 AGN pairs, respectively.}	
\label{sdss}
\end{figure}

\begin{figure*}
\centering\includegraphics[width = 18cm,height=4cm]{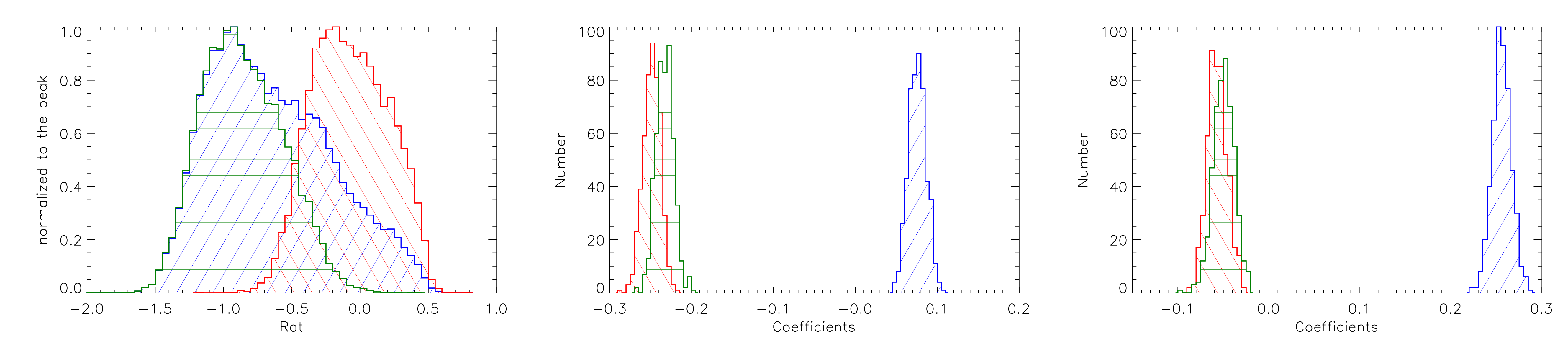}
\caption{Left panel shows the distributions of flux ratio of [O~{\sc iii}]$\lambda5007$\AA~ to narrow H$\alpha$ for the AGNs 
(histogram filled with red lines), the HII galaxies (histogram filled with dark green lines) and all the narrow emission line 
galaxies (histogram filled with blue lines). Middle panel and right panel show the determined distributions of the Spearman 
Rank correlation coefficient, after accepted the uniform distributions (with 1 as the mean value and [0.1, 10] as the minimum 
and maximum values) and the Gaussian distributions (with 1 as the mean value and 0.5 as the second moment) for the peak intensity 
ratios.  
}
\label{ran}
\end{figure*}

\begin{figure}
\centering\includegraphics[width = 8cm,height=6cm]{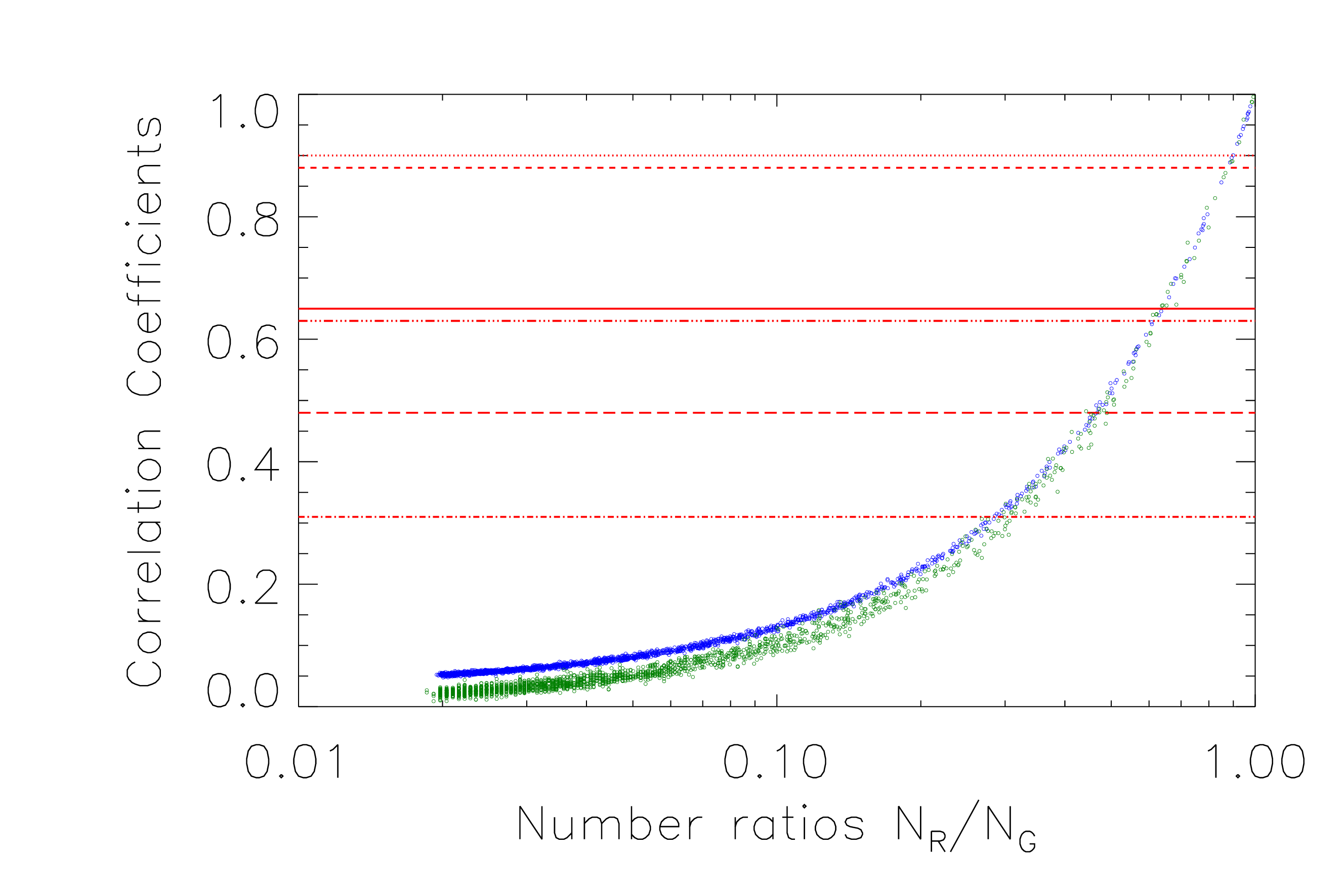}
\caption{On the dependence of Spearman Rank Correlation coefficient on the number ratio of $N_R/N_G$. Symbols in blue and in 
dark green show the results through all the galaxy pairs, and through the randomly collected 1618 galaxy pairs. Horizontal red 
solid line marks coefficient to be 0.65 for all the DPN galaxies, and horizontal red dashed lines from top to bottom show the 
coefficients to be 0.90, 0.88, 0.63, 0.48, 0.31 for the galaxies in the group of G1, G2, G4, G3, G5, respectively.
}
\label{prob}
\end{figure}

\section{Main results and necessary discussions} 

	The DPN galaxies, collected from the public sample reported in \citet{ge12}, have been classified into five groups 
through properties of narrow/broad emission lines. In \citet{ge12}, through the spectra of main galaxies in SDSS 
DR7, after removing the host galaxy contributions determined by the STARLIGHT, the emission lines are measured by multiple 
Gaussian functions. After checking the best-fitting results to the emission lines by the F-test technique, about 3030 DPN 
galaxies have been collected with apparent DPNELs with significance levels higher than 3$\sigma$. Moreover, we do not consider 
the more than 10000 galaxies collected with asymmetric or top-flat profiles of emission lines in \citet{ge12}. Meanwhile, in 
order to check the probable correlation between $R_{B}$ (flux rations of the blue-shifted components in DPNELs) and $R_{R}$ 
(flux rations of the red-shifted components in DPNELs), the 1618 of 3030 galaxies are collected from \citet{ge12} with the 
measured line fluxes of the blue/red-shifted emission components in [O~{\sc iii}] and in narrow H$\alpha$ at least five times 
larger than their corresponding uncertainties. Among the 1618 galaxies, considering the appearance of broad H$\alpha$, the 
first group (G1) includes 56 Type-1 AGNs. For the other 1562 galaxies, assumed the DGSs for the DPNELs, the 
measured emission fluxes of the blue-shifted and the red-shifted components in the DPNELs can be applied to determine the 
classifications of the two individual galaxies in the assumed DGS, through the commonly applied BPT diagrams \citep{bpt, kb01, 
ref21, kb06, ref20, ref22}. Then, the second group (G2) includes 717 galaxies with each individual galaxy classified as 
Type-2 AGN. The third group (G3) includes 260 galaxies with one individual galaxy classified as Type-2 AGN but the other 
individual galaxy classified as HII galaxy. The four group (G4) includes 492 galaxies with each individual galaxy classified 
as HII galaxy. The fifth group (G5) includes 93 galaxies with one individual galaxy not classified by BPT diagrams due to 
lack of emission components of some narrow emission lines.

	Through the reported emission line fluxes of the red-shifted and blue-shifted components in both narrow [O~{\sc iii}] 
and narrow H$\alpha$ for the 1618 galaxies included in the five groups, Fig.~\ref{rat} shows the correlations between $R_{B}$ 
(flux ratio of the blue-shifted component in [O~{\sc iii}] to the blue-shifted component in narrow H$\alpha$) and $R_{R}$ 
(flux ratio of the red-shifted component in [O~{\sc iii}] to the red-shifted component in narrow H$\alpha$). For all the 
1618 galaxies, the Spearman Rank correlation coefficient is about 0.65 ($P_{null}<10^{-21}$). Meanwhile, for the galaxies in 
G1, G2, G3, G4, G5 groups, corresponding Spearman Rank correlation coefficients are about 0.90 ($P_{null}\sim1.2\times10^{-21}$), 
0.88 ($P_{null}<10^{-21}$), 0.31 ($P_{null}\sim6\times10^{-7}$), 0.63 ($P_{null}<10^{-21}$), 0.48 ($P_{null}\sim1.5\times10^{-6}$), 
respectively.

	Furthermore, considering the uncertainties in both coordinates, through the Least Trimmed Squares regression technique 
\citep{ref23, ref24}, the linear correlations shown in Fig.~\ref{rat} can be described as 
\begin{equation}
\begin{split}
	&\log(R_{R})=0.10\pm0.01+(0.82\pm0.02)\log(R_{B}) \ \ \ (ALL)  \\
	&\log(R_{R})=0.12\pm0.03+(0.86\pm0.07)\log(R_{B}) \ \ \ (G1)  \\
	&\log(R_{R})=0.09\pm0.02+(0.78\pm0.02)\log(R_{B}) \ \ \ (G2) \\
	&\log(R_{R})=0.01\pm0.09+(0.99\pm0.16)\log(R_{B}) \ \ \ (G3) \\
	&\log(R_{R})=0.06\pm0.01+(0.86\pm0.05)\log(R_{B}) \ \ \ (G4) \\
	&\log(R_{R})=0.94\pm1.65+(1.82\pm1.71)\log(R_{B}) \ \ \ (G5) \\
\end{split}
\end{equation}
with corresponding 1RMS scatters about 0.269, 0.276, 0.266, 0.328, 0.201, and 0.436, respectively. It is clear that except the 
objects in the G5 group, the determined best fitting results and the determined Spearman Rank correlation coefficients can be 
applied to confirm the robust dependence of $R_{R}$ on $R_B$. For the dependence of $R_{R}$ on $R_B$ in the galaxies in the G5 
group, although its scatter being about two times larger than those of the dependencies of the objects in the other groups and the 
determined slope 1.82 being not apparently larger than its corresponding uncertainty 1.71, the Spearman Rank correlation about 
0.48 can be accepted to support the dependence of $R_{R}$ on $R_B$ in the objects in the G5 group. In one word, the dependence 
of $R_{R}$ on $R_B$ is common in the DPN galaxies collected from \citet{ge12}. However the linear dependence can not be expected 
by the DGSs.

	Before proceeding further, it is necessary to check whether there was flux connection between narrow emission lines in 
two nearby galaxies. In SDSS DR16 (Sloan Digital Sky Survey, Data Release 16) \citep{ref31}), there are 77982 narrow emission 
line galaxies collected with redshift smaller than 0.35, and with the measured emission line fluxes at least 5 times larger than 
their uncertainties in the [O~{\sc iii}]$\lambda5007$\AA~ and in narrow H$\alpha$, and with the median spectral signal-to-noise 
lager than 10, and with the SDSS pipeline provided subclasses as 'AGN' or 'STARFORMING' OR 'STARBURST', and with the flux ratio 
of narrow H$\alpha$ to narrow H$\beta$ smaller than 6 (to ignore effects of dust obscurations). Then, among the collected 77982 
narrow emission line galaxies, there are about 13000 Type-2 AGNs. Among the 13000 Type-2 AGNs, there are 669 couples of AGNs 
(AGN pairs) with redshift difference smaller than 0.001 and with the position distance smaller than 30arcmins. Here, if there 
were more than one AGN near to the target AGN within position distance smaller than 30arcmins, the one with the smallest position 
distance is accepted as the companion AGN to the target AGN. Then, the left panel of Fig.~\ref{sdss} shows the flux correlations 
in [O~{\sc iii}]$\lambda5007$\AA~ and in narrow H$\alpha$ in the 669 AGN pairs, leading to the corresponding Spearman rank 
correlation coefficients about -0.01 ($P_{null}\sim78\%$) and 0.06 ($P_{null}\sim13\%$) for the correlations on fluxes of 
[O~{\sc iii}]$\lambda5007$\AA~ and on fluxes of narrow H$\alpha$, respectively. The right panel of Fig.~\ref{sdss} shows the 
dependence of the flux ratios of [O~{\sc iii}]$\lambda5007$\AA~ and of narrow H$\alpha$ on the position distances in the 669 AGN 
pairs, leading to the corresponding Spearman rank correlation coefficients about 0.03 ($P_{null}\sim40\%$) and -0.002 
($P_{null}\sim97\%$) for the correlations on fluxes of [O~{\sc iii}]$\lambda5007$\AA~ and on fluxes of narrow H$\alpha$, 
respectively. Therefore, through the collected 669 AGN pairs, there are no narrow emission line flux correlations, nor dependence 
of narrow emission line flux ratios on position distances. 

	Meanwhile, among the 77982 narrow emission line galaxies in SDSS DR16, based on the same criteria on redshift difference 
smaller than 0.001 and position distance smaller than 30arcmins, there are about 3299 couples of a type-2 AGN plus a HII galaxy, 
and 11442 couples of HII galaxies. And the corresponding Spearman Rank correlation coefficients are not larger than 0.1 for the 
narrow emission line flux correlations nor for the dependencies of narrow emission line flux ratios on position distances. 
Moreover, criterion on smaller position distance smaller than 20arcmins (10arcmins) have also been applied, leading to smaller 
number of couples of narrow emission line galaxies, but leading to the similar conclusions that the corresponding Spearman Rank 
correlation coefficients are not larger than 0.1. Here, besides the results shown in Fig.~\ref{sdss} for the couples of Type-2 
AGNs, there are no plots for correlations and dependencies for the other kinds of couples of narrow emission line galaxies.

	Furthermore, as known, in order to detect apparent double-peaked features in narrow emission lines, the peak intensities 
of each DPNEL should be not very different. Therefore, it is necessary to check probable effects of peak intensity ratio on 
the results shown in Fig.~\ref{rat}. Assumed that the intensity ratio between 0.1 and 10 of two peaks in one DPNEL could lead 
to detectable double-peaked features, followed the distributions of the flux ratios $Rat$ of [O~{\sc iii}] to narrow H$\alpha$ 
of the collected narrow emission line galaxies shown in the left panel of Fig.~\ref{ran} (for AGNs, for HII galaxies, and for 
all the narrow emission line galaxies), total intensities (flux in arbitrary units) of [O~{\sc iii}] and narrow H$\alpha$ are 
artificially set to be $Rat$ and 1. Then, random values $R_3$ and $R_\alpha$ are collected from uniform distributions (with 1 
as the mean value and [0.1, 10] as the minimum and maximum values) for the intensity ratio of the red-shifted component to the 
blue-shifted component in the [O~{\sc iii}] and in the narrow H$\alpha$, leading the simulated fluxes of red-shifted and 
blue-shifted components in [O~{\sc iii}] and narrow H$\alpha$ to be $Rat\times\frac{R_3}{1+R_3}$, $Rat\times\frac{1}{1+R_3}$, 
$1\times\frac{R_\alpha}{1+R_\alpha}$ and $1\times\frac{1}{1+R_\alpha}$. Then, the corresponding Spearman Rank correlation 
coefficients can be determined between the simulated $R_{R}$ and $R_B$. Repeating the procedure above 500times, the distributions 
of the coefficients are shown in the middle panel of Fig.~\ref{ran}. Meanwhile, if random values of $R_3$ and $R_\alpha$ are 
not from uniform distributions but from Gaussian distributions (with 0 as the mean value and 0.5 as the second moment in 
logarithmic space), the corresponding distributions of correlation coefficients are shown in the right panel of Fig.~\ref{ran}. 
It is clear that the peak intensity ratios can lead to simulated Spearman Rank correlation coefficients not larger than 0.3. 
Therefore, the shown results in Fig.~\ref{rat} are not due to narrow range of peak intensity ratios.

	Once we reasonably accept the robust conclusion that there are no correlations between narrow emission line fluxes nor 
dependencies of narrow emission line flux ratios on position distances in couples of narrow emission line galaxies, the DGS is 
not the preferred scenario to explain the results in Fig.~\ref{rat}. However, it will be expected that the red-shifted components 
and the blue-shifted components in DPNELs should have similar physical dynamical environments. Therefore, the results in 
Fig.~\ref{rat} can be applied to roughly determine the proportion of DPN galaxies have their DPNELs are not related to DGSs.

	Based on all the $N_G$ (=669+3299+11442) couples of narrow line galaxies in SDSS DR16, $N_{R}$ ($\le N_G$) non repeating 
couples are randomly collected. For each selected couple, the fluxes in narrow emission lines are re-set to be equal values. 
Then, the corresponding Spearman rank correlation coefficient can be estimated through the newly set emission line fluxes in 
narrow emission lines in the $N_{G}$ couples. Repeating the procedure above about 2000 times with random values of $N_{R}$, 
Fig.~\ref{prob} shows the determined dependence of Spearman Rank correlation on the number ratio of $N_R$ to $N_G$. In order 
to find the strong linear dependency with correlation coefficient about 0.65 for all the DPN galaxies, at least 65.5\% of the 
dDPN galaxies have their DPNELs not related to the proposed DGSs. Moreover, different $N_G=1618$ (the number of the DPN 
galaxies in \citealt{ge12}) has been also applied, leading to the similar results, indicating few effects of different $N_G$ 
on the results shown in Fig.~\ref{prob}. Meanwhile, if individually considering the DPN galaxies in the G1, G2, G4, G3 and G5 
groups, the corresponding number ratios are about 91.5\%, 88.3\%, 62.5\%, 47.8\% and 30.1\%, respectively.

	Before ending the section, three additional points are noted. First, there are different Spearman Rank correlation 
coefficients for the DPN galaxies in the 5 groups, however, we have no clear points on the physical origin for the different 
coefficients and then for the very different expected number ratios of $N_R$ to $N_G$. Probably, physical dynamical environments 
have more apparent effects on the DPNELs in AGN related NLRs than on the DPNELs in HII related NLRs? However, the different 
Spearman Rank correlation coefficients and the corresponding different number ratios of $N_R$ to $N_G$ could probably indicate 
larger effects of intrinsic AGN variability on the detected DPNELs. Second, the results shown in Fig.~\ref{sdss} are only 
through the couples of nearby galaxies within 30arcmins (to 10arcmins), probably, when the merger process leading to the two 
NLRs close enough, probably some unknown physical mechanisms could lead to intrinsic connections between the two NLRs properties 
in the two galaxies undergoing the merge process. Third, based on a defined parameter $R_{ps}$ as ratio of peak 
separation to the sum of the line width of the two components in each DPNEL, we have checked whether more apparent DPNELs with 
larger $R_{ps}$ can lead to different results. Then, in each group, about half of the galaxies are collected by $R_{ps}$ larger 
than the mean value of $R_{ps}$, and shown as open red circles in Fig.~\ref{rat}. The corresponding Spearman Rank correlation 
coefficients between $R_R$ and $R_B$ are about 0.84 ($P_{null}\sim1.5\times10^{-6}$), 0.87 ($P_{null}<10^{-21}$), 0.40 
($P_{null}\sim2.7\times10^{-5}$), 0.70 ($P_{null}<10^{-21}$) and 0.44 ($P_{null}\sim8.9\times10^{-3}$) for about half of the 
objects with larger $R_{ps}$ in the group of G1, G2, G3, G4, G5, respectively, similar as the results for all the galaxies in 
each group. Therefore, there are few effects of $R_{ps}$ on our discussed results. On studying an independent large sample 
of DPN galaxies in the near future could provide further clues on understanding the discussed results.

\section{Conclusions}

\begin{itemize}
\item Through the large sample of DPNELs in SDSS, strong linear correlations can be found between $R_R$ and $R_B$.
\item Through the collected galaxy pairs with redshift difference smaller than 0.01 and position distance smaller than 30arcmins, 
	there are no flux connections between narrow emission lines in the different kinds of galaxy pairs.
\item Considering narrow range of peak intensity ratios for detecting DPNELs, the estimated correlation coefficients are 
	smaller than the coefficients for the DPNELs in SDSS.
\item The strong linear correlations between $R_R$ and $R_B$ are robust enough, strongly indicating that the commonly proposed 
	dual galaxy system is not the preferred scenario to explain the reported DPNELs.
\item Through simple simulating results on real flux ratio distributions of [O~{\sc iii}] to narrow H$\alpha$, more than 
	65.5\% of the DPNELs are not related to DGSs. 
\item Considering the DPN galaxies in G1, G2, G4, G3 and G5 groups, more than 91.5\%, 88.3\%, 62.5\%, 47.8\% 
	and 30.1\% of the DPNELs are not related to DGSs, respectively.
\end{itemize}

\begin{acknowledgements}
Zhang gratefully acknowledges the anonymous referee for giving us constructive comments and suggestions to greatly improve the 
paper. Zhang gratefully thanks the kind financial support from GuangXi University and the kind grant support from NSFC-12173020 
and NSFC-12373014 and the Guangxi Talent Programme (Highland of Innovation Talents). This manuscript has made use of the data 
from the SDSS projects. The SDSS-III web site is http://www.sdss3.org/. SDSS-III is managed by the Astrophysical Research 
Consortium for the Participating Institutions of the SDSS-III Collaboration.
\end{acknowledgements}

\end{document}